\documentclass[10pt,a4paper]{article}
\usepackage{graphicx}
\usepackage{color}
\usepackage{siunitx}
\usepackage{xspace} 
\usepackage{authblk}

\usepackage{pdflscape}

\newcommand{\fluoro}{$^{18}$F\xspace}
\newcommand{\pterf}{para-terphenyl\xspace}
\newcommand{\mc}{MC\xspace}

\begin{document}
\title{\textbf{Feasibility of the $\beta^-$  Radio-Guided Surgery with a Variety of Radio-Nuclides of Interest to Nuclear Medicine}}

\author[1,2] {Carlo~Mancini-Terracciano}
\author[2,3]{Raffaella~Donnarumma\footnote{co-first author} }
\author[4]{Gaia~Bencivenga}
\author[2]{Valerio~Bocci}
\author[5]{Antonella~Cartoni}
\author[1,2]{Francesco~Collamati} 
\author[5]{Ilaria~Fratoddi}
\author[6]{Alessandro~Giordano}
\author[7]{Luca~Indovina}
\author[8]{Michela~Marafini}
\author[2]{Silvio~Morganti}
\author[9]{Dante~Rotili} 
\author[2,3,10]{Andrea~Russomando}
\author[11]{Teresa~Scotognella}
\author[2,3]{Elena~Solfaroli~Camillocci}
\author[12]{Marco~Toppi}
\author[2,3]{Giacomo~Traini}
\author[5]{Iole~Venditti}
\author[2,3]{Riccardo~Faccini}

\affil[1]{Dip. Scienze di Base e Applicate per l'Ingegneria, Sapienza Univ. di Roma, Rome, Italy;}
\affil[2]{INFN Sezione di Roma, Rome, Italy;}
\affil[3]{Dip. Fisica, Sapienza Univ. di Roma, Rome, Italy;}
\affil[4]{PET-CT Center,  Policlinico A. Gemelli, Rome;}
\affil[5]{Dip. Chimica, Sapienza Univ. di Roma, Rome, Italy;}
\affil[6]{Ist. Medicina Nucleare, Univ. Cattolica del Sacro Cuore, Rome, Italy;}
\affil[7]{UOC Fisica Sanitaria,  Policlinico A. Gemelli, Rome;}
\affil[8]{Museo Storico della Fisica e Centro Studi e Ricerche ``E. Fermi'', Rome, Italy;}
\affil[9]{Dip. Chimica e Tecnologie del Farmaco, Sapienza Univ. di Roma, Rome, Italy;}
\affil[10]{Center for Life Nano Science@Sapienza, Istituto Italiano di Tecnologia, Rome, Italy.}
\affil[11]{Ist. Medicina Nucleare,  Policlinico A. Gemelli, Rome;}
\affil[12]{Laboratori Nazionali di Frascati INFN, Frascati, Italy;}

\maketitle
\pagestyle{empty}
\thispagestyle{empty}

\begin{abstract}

The $\beta^-$ based radio-guided surgery overcomes the corresponding $\gamma$ technique in case the background from healthy tissues is relevant. It can be used only in case a radio-tracer marked with $^{90}$Y  is available since the current probe prototype was optimized for the emission spectrum of this radio-nuclide. Here we study, with a set of laboratory tests and simulations, the prototype capability in case a different radio-nuclide is chosen among those used in nuclear medicine.

As a result we estimate the probe efficiency on electrons and photons as a function of energy and we evaluate the feasibility of a radio-guided surgery exploiting the selected radio-nuclides. We conclude that requiring a 0.1~ml residue to be detected within 1~s by administering 3~MBq/Kg of radio-isotope, the current probe prototype would  yield a significant signal in a vast range of values of SUV and TNR in case $^{31}$Si,$^{32}$P, $^{97}$Zr, and $^{188}$Re are used. Conversely, a tuning of the detector would be needed to efficiency use $^{83}$Br,  $^{133}$I, and $^{153}$Sm, although they could already be used in case of high SUV or TNR values. Finally, $^{18}$F,$^{67}$Cu, $^{131}$I,  and $^{177}$Lu are not useable for radio-guided surgery with the current  probe design.
\end{abstract}



\section*{Introduction}
Radio-guided surgery (RGS)  is a technique that enables the surgeon to evaluate in real time the completeness of the tumor lesion resection. At the same time RGS allows to minimise the amount of healthy tissue removed. It represents a significant adjunct to intra-operative detection of millimetre-sized tumor residues, providing the surgeon with vital and real-time information regarding the location and the extent of the lesion. It is crucial for those tumors where surgical resection is the only possible therapy, since it reduces the probability of tumor recurrence assessing surgical resection margins. 

RGS makes use of a radio-labelled tracer and of a probe. The radiopharmaceutical is administered to the patient before surgery and  its uptake has to be higher in the tumor than in healthy organs, allowing to discriminate among them.

Traditional RGS makes use of $\gamma$ emitters as tracer~\cite{GammaRGS,GammaProbes,GammaReview}. 
However, the use of this technique is strongly limited, and often even prevented, by the   background produced from the uptakes of the healthy tissue. Such a background is non-negligible because  $\gamma$ radiation can travel through large amounts of tissue. Moreover, the large mean free path of $\gamma$ radiation exposes the medical personnel to radiation risks.
\begin{table}[!tbh]
\centering
\scriptsize
\begin{tabular}{l*{5}{c}}
\hline
Isotope &  $\tau_{1/2} (h) $   & $E_{\gamma}$ (keV) &  $I_\gamma$ (\%) & $ EP_{\beta}$ (keV) & $I_\beta$ (\%) \\
\hline
$^{31}$Si  & 2.62	     &&   &  1491        &  100  \\            
 $^{32}$P  &   343	     &&  &  1710         &   100 \\            
 $^{67}$Cu  &	62     &93& 16  &      377    &   57 \\            
  &	       &  184 &48   & 468     & 22    \\            
  &	       &   &   & 561     & 20    \\            
 $^{83}$Br  &	2.4     && &     935     &99    \\            
 $^{90}$Y  &	64     & &   &     2280     &100    \\  
 $^{97}$Zr  &	17     &743& 93 &     759     &88    \\  
(secondary*)  &	     &657& 86 &    1277     &86    \\  
 $^{131}$I  &	192     & 365&82  &334          &  7  \\            
            &	        &637&7  &   606       & 90   \\            
 $^{133}$I  &	20.8     & 530&87  &1227          &  83.4  \\            
$^{153}$Sm  &	46     & 103&29  &635          &  31  \\            
            &	        &&  &   704       & 49   \\            
            &	        &&  &   808       & 18   \\            
 $^{177}$Lu  &	160     &112& 6 &     177     & 12   \\            
             &	        &208 & 10 &   500       & 79   \\            
$^{188}$Re  &	 17    & 155 & 15  & 1962         & 25    \\            
  &	       &   &   & 2118         & 72    \\            
 \hline
\end{tabular}
\caption{Characteristics of the $\beta^-$ radio-isotopes of interest for the proposed RGS technique:  half-life ($\tau_{1/2}$), $\gamma$ and $\beta^-$  intensities ($I_\gamma$, $I_{\beta}$), photon lines energy (E$_\gamma$), and $\beta^-$ end-points ($EP_\beta$). Only components with $I>5\%$ are listed.(*) the second line of  $^{97}$Zr is due to the subsequent decay, with $\tau_{1/2}=72$~min of the  $^{97}$Nb produced in the primary decay.}
\label{Tab:beta-}
\end{table}

The use of $\beta^-$ radiation to overcome the limitations of traditional RGS was recently proposed~\cite{SciRep,PatentBeta-RGS}. Indeed, electrons have a very small penetration. Therefore a $\beta^-$ RGS has a more favourable  ratio between the signal coming from the tumor and the rest of the body. An ex-vivo test has shown the feasibility of this technique for a meningioma case~\cite{Paziente0}. Meningioma has been chosen because it is known to express receptors to DOTATOC~\cite{MeningiomaUptake}, a somatostatin analogue that could be labelled with  $^{90}$Y, a pure $\beta^-$ emitter with a 2.3~MeV electron energy endpoint. 

Limiting to such radio-tracer, though, would mean restricting this technique to very few tumors, mainly glioma~\cite{MeningiomaRGS} and neuro-endocrine tumors~\cite{RENET}. To further broaden the applicability of the technique, the use of different radio-isotopes, with larger gamma contamination and lower endpoint is under investigation. 

Among the $\beta^-$-emitting radio-isotopes that have the largest diffusion in nuclear therapy~\cite{RT-tracers}, we considered those that have an electron energy end-point
of at least 500 keV, dominant $\gamma$ lines below 300 keV and total $\gamma$ intensities below 100\%: $^{32}$P, $^{131}$I, $^{133}$I, $^{153}$Sm, $^{177}$Lu, and $^{188}$Re. In addition we  considered isotopes that are chemically in the same family as nuclides already used to mark tracers since these are the best candidates to give life to a new radio-tracer. We thus included in our studies $^{31}$Si, $^{67}$Cu, $^{83}$Br, and $^{97}$Zr. The characteristics of the decays of the considered isotopes are detailed  in Tab.~\ref{Tab:beta-}.

\begin{figure*}[!bht]
\centering
  \includegraphics [width=0.435\textwidth]{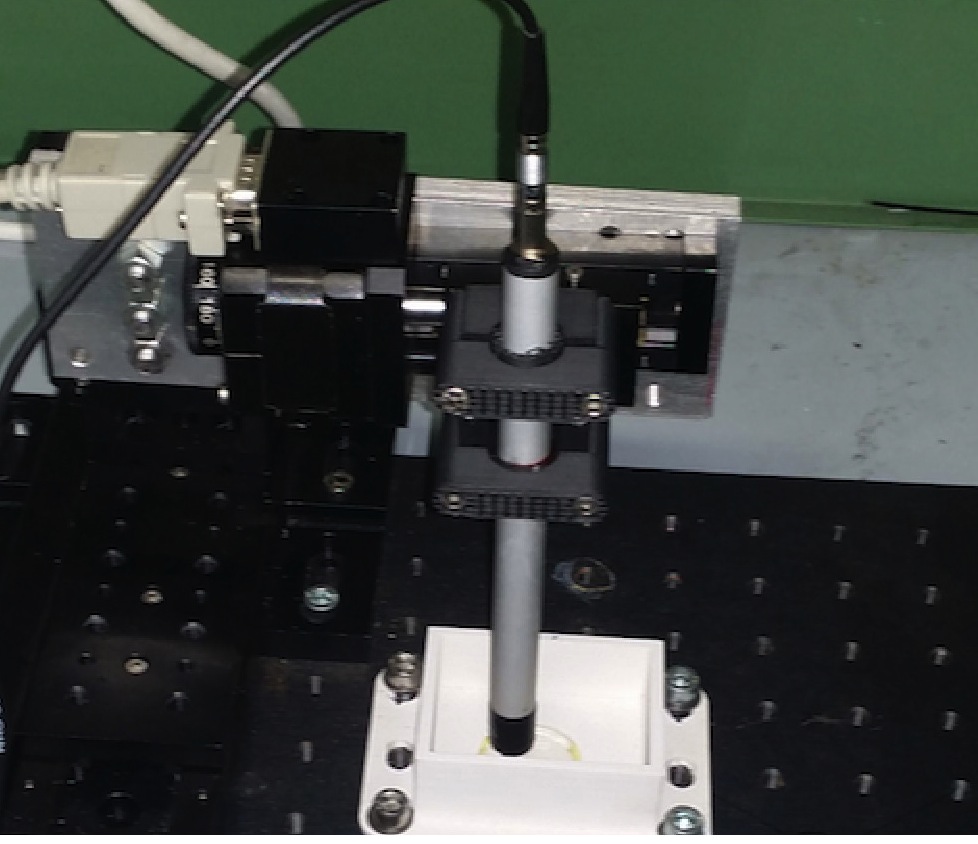}
 \includegraphics [width=0.48\textwidth]{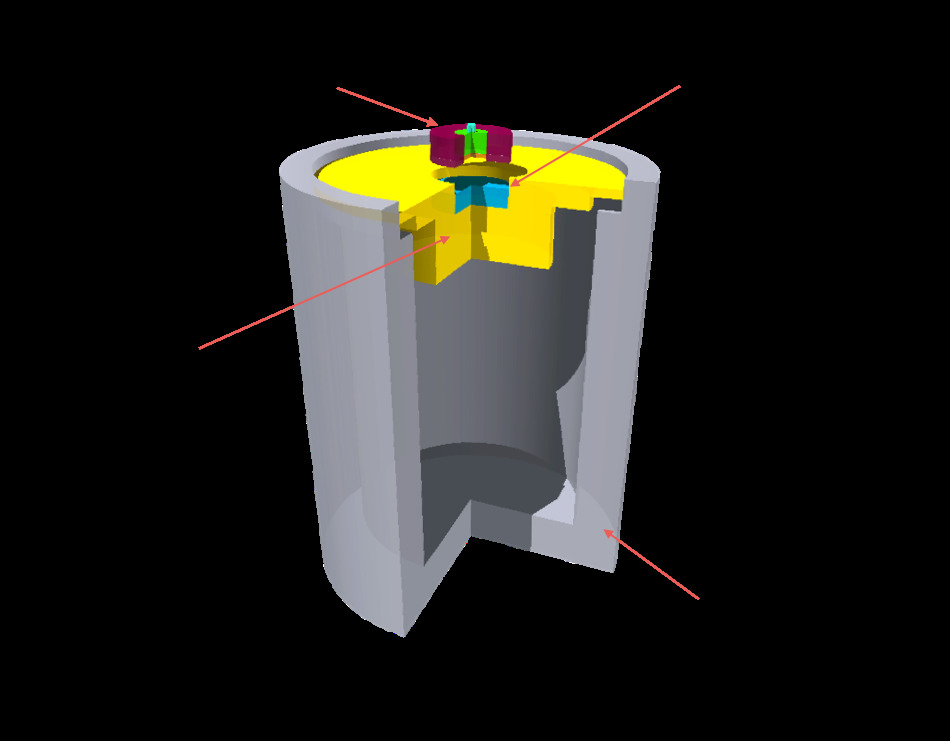}
\caption{The probe prototype and the experimental setup of the measurents with the sealed point-like sources (left) and scheme of the setup with \fluoro (right).}
 \label{fig:setup}
 \end{figure*}

Another alternative is to consider \fluoro, which has $\tau_{1/2}=1.8$~h and a $\beta^+$ emission with end-point at 633.5~keV and 97\% intensity and is used to radio-label the fluorodeoxyglucose (FDG). The uptake of the latter is the same as glucose, which in turn is correlated with tissue metabolism. It is widely used as  marker for the positron emission tomography (PET) and, therefore, if the developed probe were sensitive to \fluoro with low background there would be a significant extension of the applicability of the technique. It is to be noted that in this case the probe would detect the positrons before they stop or annihilate, while annihilation photons would represent a background. The possibility to intra-operatively detect $\beta^+$ decays has already been explored in the past~\cite{beta+technique, beta+preclinical} with the development of ad-hoc dual detectors for the simultaneous measurement of positrons and $\gamma$s. The difficulties in developing and handling the probe has limited the clinical studies~\cite{beta+clinical}.

This paper describes the laboratory tests performed to measure the probe efficiency to $\gamma$ and  $\beta$ radiations, studies that are the basis for the extention of the technique to radio-isotopes other than the baseline $^{90}$Y. To investigate potentialities of radiations significantly different, several sources with different $\beta$ spectra and gamma lines were used.

After tuning the simulation on these measurements, the expected signal from tumor and background assuming to use the different possible radionuclides of Tab.~\ref{Tab:beta-} was estimated. Finally in the comparison among radio-isotopes we also estimated the expected dose rate delivered to the medical personnel during surgery. This requirement, in addition to the need to minimize the dose to the patient, limits the activity that can be administered.

\section*{Material and Methods}

\subsection*{The Probe Prototype}

The probe prototype considered in this paper is visible in Fig.~\ref{fig:setup}. 
The lower black part in the picture is a ring of \SI{12}{mm} external diameter in Acrilonitrile Butadiene Stirene (ABS) that shields the sensitive element content in it from the radiation coming from the side.
The sensitive part of the probe is a \SI{3}{mm} height and \SI{5}{mm} diameter 
 of  mono-crystalline \pterf  doped to 0.1\% in mass with diphenylbutadiene. This plastic scintillator was adopted because of its high light yield ($\sim$3 times larger than typical organic scintillators), non-hygroscopic property, and low density~\cite{PTerf} that minimises the sensitivity to photons. 
The light tightness around the scintillator is ensured by a \SI{15}{\micro\meter} thick aluminum sheet. 
The  scintillation light is read out by  a SiPM (sensL B-series 10035) biased with \SI{24.5}{V}. Its spectral range is $300-\SI{800}{nm}$ and the peak wavelength is \SI{420}{nm}.
An aluminum cylindrical body (diameter \SI{12}{mm} and length \SI{14}{cm}) encapsulates this assembly. A portable electronics based on ArduSiPM, with ethernet connection to a PC was used for the read out~\cite{ardusipm}. 

\subsection*{Experimental Setups}

We measured the probe sensitivity to the radiation emitted by several long life-time radioisotopes with different energies, contained in sealed sources and  in liquid \fluoro samples. They contain radio-isotopes by definition different from the ones of interest to RGS, which requires half-lives shorter than few days. Therefore  these measurements can only been used to optimize the parameters used in the \mc simulations (see Sec.~\ref{sec:FLUKA}) to describe the probe behavior, in particular the optical properties of \pterf. The \mc is  in turn used to estimate the RGS performances.

This measurement campaign has been performed with two different setups: the setup used for the sealed laboratory sources and the one made to measure the probe efficiency to liquid \fluoro. These two setups will be described separately in the next subsections.
\begin{table*}[!tbh]
\centering
\scriptsize
\begin{tabular}{l*{10}{c}}
\hline
Isotope & A (kBq)&  $\tau_{1/2} $ (y)  & $E_{\gamma}$ (keV) &  I (\%) & $ EP_{\beta^-}$ (keV) & I (\%)& $EP_{\beta^+}$(keV) &  I (\%) & $e^-$ (keV)    &  I (\%)   \\
\hline
 $^{60}$Co  & 36.7$\pm$ 1.1 &	 5.3    &1173& 100 & 317          & 100        &               &                &                 &                \\
    &                 &	          & 1332 & 100  &                &               &                &                &                 &                \\
 $^{90}$Sr  & 3.3$\pm$ 0.1 &	 28.9    &&  & 540          & 100        &               &                &                 &                \\
    &                 &	          &   &    & 2280           & 100               &                &                &                 &                \\
$^{22}$Na  & 40.5$\pm$ 1.2  &	 2.6    & 511 & 181 &               &               & 546          & 90           &                 &               \\
                 &    &	          &1274&100 &               &                &               &                &                 &               \\
$^{152}$Eu  & 41.5$\pm$ 1.2 &  13.2  & 40   & 58  & 384         & 2.4           &                &               &   75          &  19          \\
               &      &	          & 45   & 11  & 694          &    14        &                &                &   114         &  11           \\
                     &	          & 122 & 28  & 1475       &     8          &                &                &   120        &  2.4          \\
&                     &	          & 244 & 8    &               &                &                &                &                 &                \\
 &                    &	          & 344 & 27  &               &                &               &                &                 &                \\
  &                   &	          & 443 & 3    &               &                &                &                &                 &                \\
   &                  &	          & 779 & 13  &               &                &                &                &                 &                \\                    
    &                 &	          & 867 & 4    &               &                &                &                &                 &                \\                     
     &                &	          & 964 & 15  &               &                &                &                &                 &                \\
      &               &	          & 1085 & 12 &               &                &                &                &                 &                \\
       &              &	          & 1112 & 4  &              &               &                &                &                 &                \\                    
        &             &	          & 1408 & 21 &               &                &                &                &                 &                \\                                              
\hline
\end{tabular}
\caption{Characteristics of the sealed sources used in the tests: activity at the moment of the experiment (A), half-life ($\tau_{1/2}$), intensities (I), photon lines energy (E$_\gamma$), and end-points (EP).}
\label{tab:sources}
\end{table*}

\subsubsection*{Sealed sources setup}
\label{sec:sealed}
The sealed sources were chosen in order to test the efficiency to photons and electrons at several energies. Their relevant characteristics are listed in Tab.~\ref{tab:sources}. The $\gamma$ sources have also $\beta$ decays from internal conversion. Since the amount of electrons exiting the source depends strongly on the its details, we added a 2.8~mm copper shielding.

All the sealed sources were point-like  a part the $^{90}$Sr one that had a diameter of \SI{16}{mm} . The probe was placed vertically on top of each of them, \SI{3.5}{mm} far from the sources. A picture of the experimental setup is in the left of Fig. \ref{fig:setup}. A bi-dimensional horizontal scan was performed for each measurement in order to center the probe with respect to the source.

Finally, to study different electron spectra, the measurements with the $^{90}$Sr were repeated also with a plastic shielding made of up to 11 sheets of 2-methyl-1,3-butadiene, 400~$\mu$m thick.

\subsubsection*{\fluoro setup}
\label{sec:fluoro}
To measure the probe efficiency to \fluoro,  \SI{0.48}{ml} of water with \SI{53}{kBq} of such isotope have been placed in a plastic (PVC) vessel, its central water receptacle was cylindrically shaped, with a diameter of \SI{12.2}{mm}. 
To shield from $\gamma$ radiation, the plastic vessel has been placed in a lead vial, also cylindrically shaped. The lead vial was much larger than the plastic vessel to reduce the background due to the annihilation $\gamma$'s. In detail: the vial internal radius was \SI{2}{cm} and the distance from the bottom \SI{4.4}{cm}. A scheme of the experimental setup, with the most relevant sizes of the lead vial shield and the vessel, is shown in Fig. \ref{fig:setup} on the left.

\subsection*{The Probe Simulation and its Optimization}
\label{sec:FLUKA}
The estimate of the expected performances of the $\beta^-$ probe prototype in case of use of $\beta^+$ radio-tracers for intra-operatorial use requires an estimate of the signal from the tumor and of the background coming from the uptake of radio-tracer from the rest of the body. These measurements need a simulation properly reproducing the measurements on the sources.

Furthermore, this study aims to quantify the efficiency on electrons and photons as a function of their energies and this also requires a detailed simulation.

To this aim, the  FLUKA Monte Carlo (MC) code~\cite{FLUKA} was used for a full simulation of the probe and of the experimental setups. From the geometry of the setup and of the probe the program  estimates the energy released in the interaction of  the decay products of the radio-nuclides with the detector  and produces 28000 optical photons per \si{MeV} of deposited energy~\cite{PTerf}. To increase the simulation precision, the $\delta$ rays  production threshold and the electron transport threshold have been set to \SI{10}{keV} in the scintillator and in the material in contact with it. Moreover, in these materials the electrons step size has been limited to make them lose no more than 1\% of their energy per step. This allows to accurately reproduce the optical photons generation points and the $\delta$ rays production at the interface with the scintillator. In all the other materials the transport threshold of all the particles and the $\delta$ ray production threshold are set at \SI{100}{keV}. 
The path of the optical photons is traced within the probe and the number of optical photons reaching the light detector (SiPM) is counted ($N_{op}$). The fraction of the decaying nuclei that release a signal is computed as the fraction of simulated decays that has $N_{op}$ greater than a threshold ($N_{th}$) representing the minimum signal detectable by the probe prototype. Such efficiency ($\epsilon^{MC}$) depends therefore on this threshold.

The simulation has several parameters which need to be set empirically. The ones that are most relevant for this analysis are related to the optical properties of the \pterf, since they determine the conversion between the released energy and $N_{op}$. While the attenuation length has been measured to be $\approx$2~cm, the diffusion coefficient ($D$ expressed in cm$^{-1}$) and the reflectivity ($r$) of the aluminum-wrapping were varied to find the configuration that best reproduces the measurements. Finally also $N_{th}$, related to the electronic threshold applied at readout, is optimized on data.

The optimization of these parameters is based on the measurement of the maximum observed rates in the scans over  $N_s-1$ sealed sources described in Sec.~\ref{sec:sealed}\footnote{of the described setups we used in the minimization only part of the measurements with $^{90}$Sr, both to avoid giving too much wight on the electron detection and to have a control sample (see right of Fig.~\ref{fig:scans}).}
 and the \fluoro setup described in Sec.~\ref{sec:fluoro}: $R_i$, $i=1,...,N_S$. Such measurements are divided by the corresponding source activity ($A_i$), to estimate the efficiency on data, $\epsilon_i^{DT}=R_i/A_i$, with an error estimated as the sum in quadrature of the statistical error on $R_i$ and the uncertainty on $A_i$ (see Tab.~\ref{tab:sources}).  

At the same time, for each source the MC estimates the efficiency for each possible value of the free parameters ($\overrightarrow{P}=\{D,r, N_{th}\}$): $\epsilon^{MC}_i(\overrightarrow{P})$. The best estimates of the parameters are those that minimize 
\begin{equation}
\chi^2(\overrightarrow{P})=\sum_i \left( \frac{\epsilon^{MC}_i(\overrightarrow{P})-\epsilon^{DT}_i}{\sigma_i} \right) ^2 \; .
\end{equation}

\subsection*{Estimate of expected RGS performances}
\label{sec:RGSsimu}
In order to quantify the performances of the probe on different radio-isotopes, a benchmark surgical configuration was established. 
We considered that a typical patient of 70 kg of weight could be administered $\approx 210$ MBq, and that a typical SUV of interest could range between 1 and 8, with a ratio between the uptake in the tumor and in the nearby healthy tissue (TNR) ranging between 4 and 30, as  established in our previous studies on cerebral tumors~\cite{MeningiomaRGS}.
\begin{figure}[!tbh]
\centering
\includegraphics [width=0.48\textwidth]{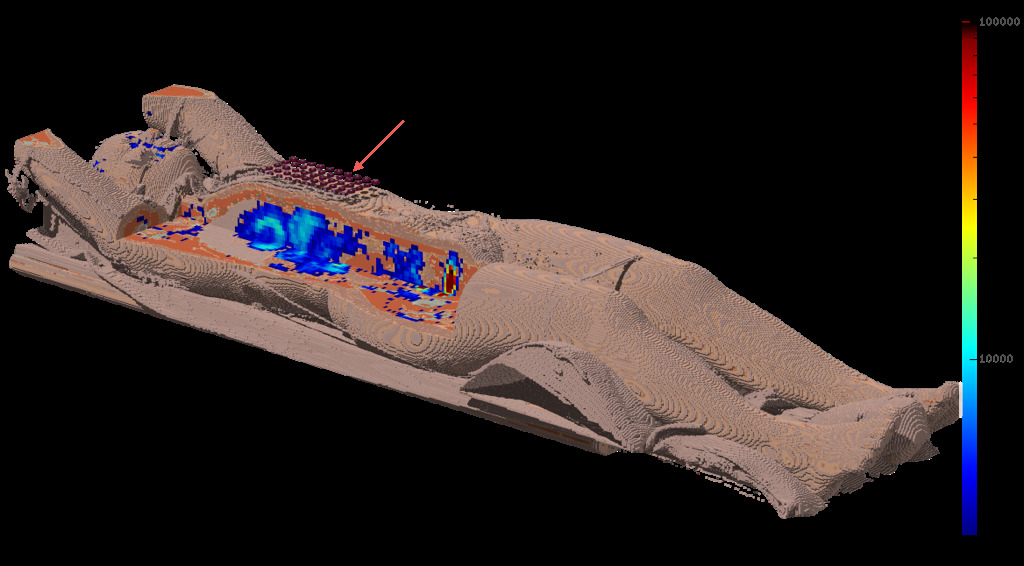}
 \caption{Geometry of the simulation used to estimate the far background}
 \label{fig:melanix}
 \end{figure}
 
 To estimate the discrimination power between tumor and healthy tissues, we performed, for each radio-isotope, two sets of \mc simulations: one to evaluate the signal from a tumor residue and the background due to the radiation emitted from the nearby healthy tissue; the other one to evaluate the background, mostly due to $\gamma$ emission resulting from the uptake in the whole body. The acquired information was then combined to estimate the minimum time needed to identify a tumor residue or the minimum specific activity that needs to be administered to have a timely response from the probe.

\subsubsection*{Signal From Tumor and Nearby Healthy Tissues}

To estimate the signal that we would measure from a small residue embedded in healthy tissue, we simulated a 6~mm diameter, 7~mm height cylinder with activity $A_{T}=A_{M}*SUV$ (where $A_M$ is the mass specific administered activity), representing the residue, encapsulated in the center of a 2~cm diameter, 1~cm height cylinder with activity $A_{B}=A_{T}/TNR$, representing the healthy tissue. The estimated rate is $R_{Tum}$.

Conversely, to estimate the signal we would have on the nearby healthy tissue, a 2~cm diameter, 1~cm height cylinder completely active with $A_{B}$ was simulated. The estimated rate is $R_{Near}$.

  \subsubsection*{Far Background}
The simulation was performed on a sample PET/CT scan provided by the Osirix  web site\footnote{ the DICOM file called ``MELANIX'' on http://www.osirix-viewer.com/datasets/} (see Fig.~\ref{fig:melanix}).  The CT voxels, have been converted from  Hounsfield Units (HU) into density and elemental composition for MC simulation\cite{Schneider:2000dn}.
The spatial distribution of the isotopes is sampled using the FDG-PET scan of the same patient, thus assuming for the hypothetical radio-tracer the same uptake as the FDG.  

To increase the simulation efficiency, and to average  on different possible positions of the probe with respect to the patient body, 50 replicas of the probe were simulated and the average counting rate ($R_{Far}$), scaled for the total activity in the patient, was considered as background rate. The probes were conservatively placed on the abdomen of the patient.

To avoid double counting the background from nearby healthy tissues, in the estimate of the far background contributions from decays in a fiducial volume around the probe were discarded. Eventually, the signal expected when the probe is in contact with a tumor residue ($S_{res}$) or healthy tissue ($S_{HT}$) are estimated as:
\begin{eqnarray}
S_{Res}&=&R_{Tum}+R_{Far}\\
S_{HT}&=&R_{Near}+R_{Far}
\end{eqnarray}

\subsubsection*{Minimum Time Needed and Administered Activity}
The benchmark of the feasibility of the use of a given radiation is the activity that needs to be administered to have a significant signal in a reasonable time. We consider to be able to efficiently detect residues if the probability of 
false negatives FN$<5\%$ and of false positives FP$<1\%$\footnote{FN and FP are computed from $S_{Res}$ and $S_{HT}$  as described in Ref.~\cite{MeningiomaRGS}  }. Under these conditions, as described in Ref.~\cite{MeningiomaRGS} we can calculate $t^{min}_{probe}$, the time needed to have such a response given the administered activity  and  $A^{min}_{1s}$, the mass specific activity needed to have a reasonable time response of $t^{min}_{probe}=1$~s.

\subsection*{Estimate of Medical Staff Exposure}
Another critical information to establish the feasibility of an RGS technique is the dose received by the medical personnel. Given that a reasonable limit on the dose received by the medical personnel is 1~mSv/year, the dose rate induced by the activity administered to the patient defines the maximum number of hours that the surgical staff can operate adopting the RGS. 

Given the limited penetration of the $\beta^-$ radiation we consider that the dominant contribution to the dose to the surgical staff comes
photons. Therefore, we estimate the expected effective dose rate from the specific gamma-ray dose rate ($\Gamma$) calculations in Ref.~\cite{specific-gamma}. In the case of pure $\beta^-$ decays we consider the effective dose delivered by the Brehmsstrahlung radiation as computed in Ref.~\cite{gamma-Y90}.
From the estimated $\Gamma$ values we extract $DR_{med}$, the expected dose rate on the medical staff in case 3MBq/kg of radio-isotope is administered to a 70 kg patient and the person is on average at a distance of 1~m. The estimated values are reported in Tab.~\ref{tab:results}.
\begin{figure*}[!bth]
\centering
\includegraphics [width=0.9\textwidth]{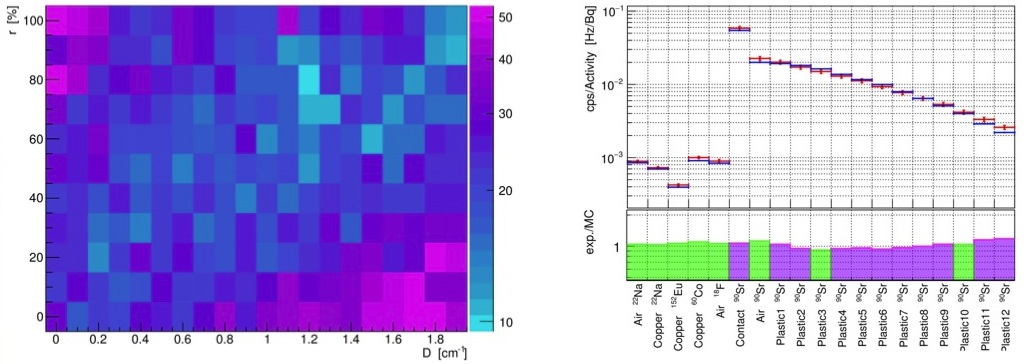}
 \caption{ Left: $\chi^2$ as a function of the diffusion coefficient ($D$) and the reflectivity ($r$) after the optimization of $N_{th}$.Right: Comparison between experimental counts and MC for the optimized values of $D$ and $r$. The residual discrepancy (exp/MC) is also shown. Green bins in the bottom panel indicate the sources included in the $\chi^{2}$ minimization.}
 \label{fig:scans}
 \end{figure*}

\section*{Results}
\label{sec:results}
\subsection*{Simulation Parameter Optimization}
The optimization of the MC parameters, the diffusion coefficient $D$, the reflectivity ($r$) and the equivalent threshold in optical photons ($N_{th}$), were determined as described in Sec. ~\ref{sec:FLUKA}.

The measured $\chi^{2}$ as a function of $D$ and $r$ is shown in Fig. \ref{fig:scans} on the left.
There is a clear correlation between the two parameters, a lower reflectivity being compensated by a lower diffusion parameter. Nonetheless there is a well defined minimum in r=80\% and D=\SI{1.2}{\per\centi\metre} and these are the adopted parameters in the remaining of the paper.

Fig.~\ref{fig:scans} on the right shows the resulting level of agreement of the measured and estimated rates together with the fractional discrepancy between the two.


\subsection*{Efficiencies}

\begin{figure}[!tbh]
\centering
 \includegraphics [width=0.48\textwidth]{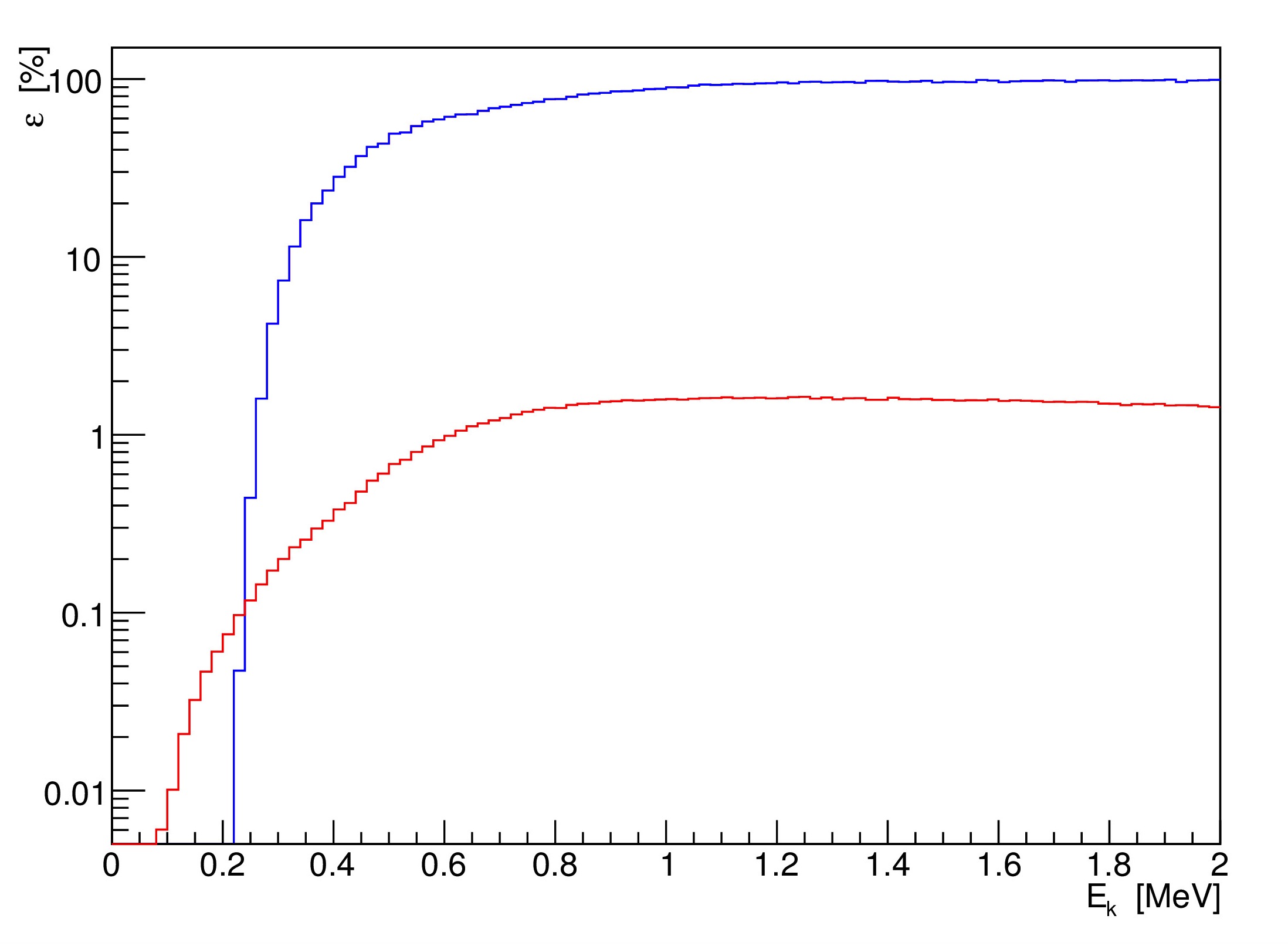}
 \caption{Probe efficiency as a function of the electron (top) and photon (bottom) energy.}
 \label{fig:eff}
 \end{figure}

After fixing the MC parameters, the efficiency of the probe on electrons and photons  could be estimated as a function of the energy. The result is reported in Fig.~\ref{fig:eff}: the efficiency on electrons reaches a plateau of 95\%, the flexus of the distribution being $\approx$400keV. Similarly, the photon efficiency reaches 1.2\% at $\approx 800$~keV and has a less steep raise.

These observations confirm that the probe is optimal for $^{90}$Y and that it has low sensitivity to photons. The performances are also significantly better than the previous prototype with PMT readout~\cite{ProbeTNS}. 

\subsection*{RGS Feasibility}

In order to compare the expected performances of the current probe prototype among radio-isotope, the simulation has been run 
as described in Sec.~\ref{sec:RGSsimu}, assuming a 70 kg patient with SUV ranging between 1 and 8 and TNR between 4 and 30.
Tab.~\ref{tab:results} reports the relevant information for a particular case, with a reasonable SUV (4) and TNR (8). As a reference also the results obtained by the same analysis with the golden radio-isotope, $^{90}$Y, are reported.

\begin{figure}[!tbh]
\centering
\includegraphics [width=0.48\textwidth]{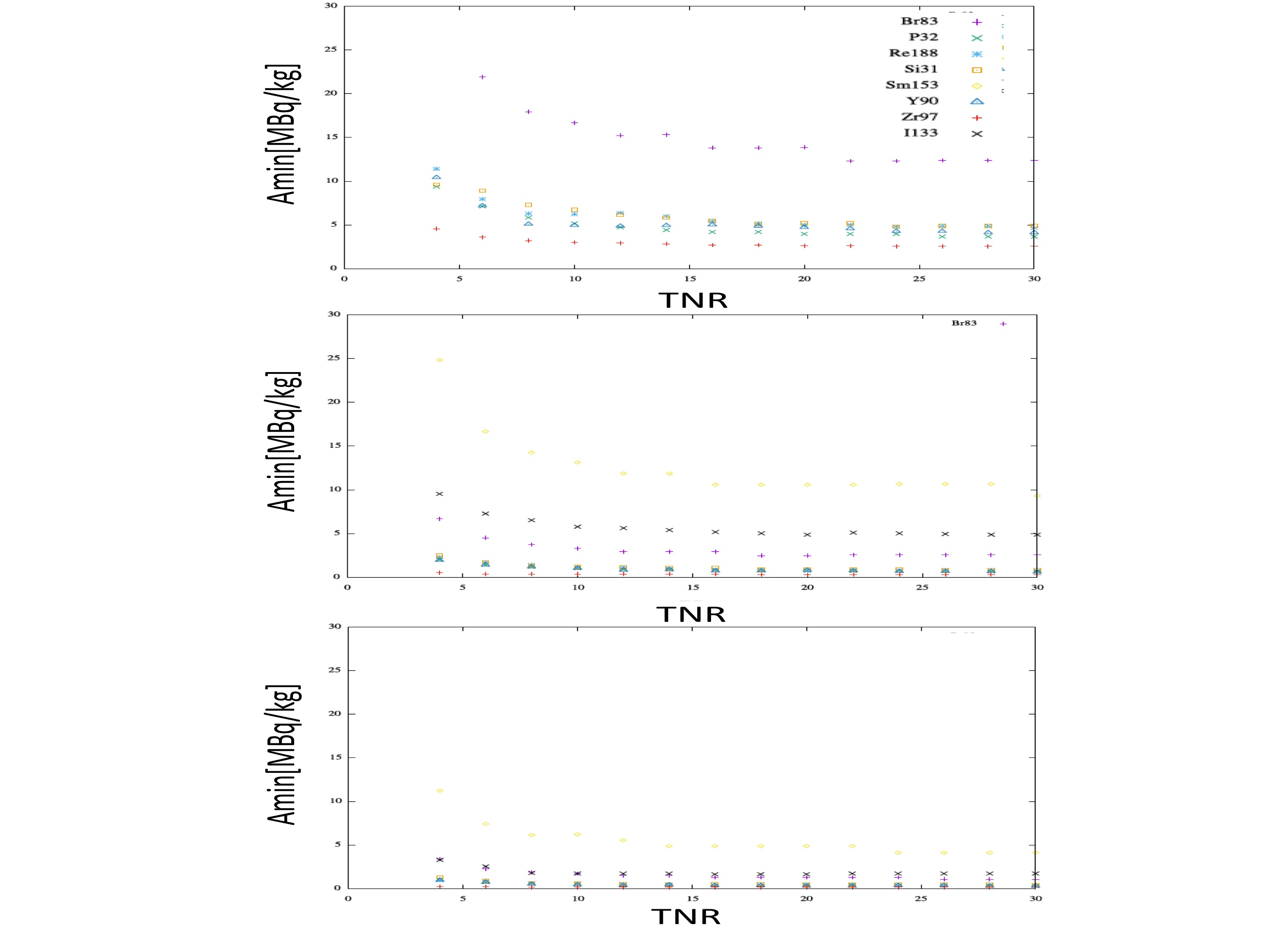}
 \caption{Minimum mass specific activity needed to have a discriminating response from the probe in less than 1~s as a function of TNR for SUV=1 (top) and SUV=4(center) and SUV=8 (bottom).  In the left plot all the points corresponding to $^{133}$I and  $^{153}$Sm are above the maximum in the axis.}
 \label{fig:results}
 \end{figure}

The extrapolation to other values of SUV and TNR are shown in Fig.~\ref{fig:results}.
\begin{table}[!bth]
 \begin{center}
  \caption{Comparison between radio-isotopes: the expected signal on a residue ($S_{res}$) and on healthy tissues ($S_{HT}$) and  the contribution from the far background ($R_{far}$); the minimum time required to identify a residue as defined in the text($t_{probe}^{min}$); the dose rate to the medical personnel($DR_{med}$) if 3~MBq/kg are administered and the average distance from the patient is 1~m; and the minimum mass specific activity ($A^{min}_{1s}$) needed to have $t_{probe}^{min}=1$~s,  These estimates are obtained assuming SUV=4, TNR=8. 
  \label{tab:results}}
  \begin{tabular}{| l || c | c |c | c | c|c||}
    \hline
     Isotope&   $S_{res}$ & $S_{HT}$ & $R_{Far}$& $t_{probe}^{min}$  & $A^{min}_{1s}$ &$DR_{med}$\\ 
 & (cps) &  (cps) &  (cps) &  (s) & (MBq/kg) & ($\mu$Sv/hr)\\  \hline
$^{18}$F 		& 	38	&	35	&	33	&	$>$25	&$>$75	& 38 \\
$^{31}$Si  	&	38	&	10	&	2.3	&	0.4		&1.3		& 0.027 \\            
 $^{32}$P  	& 	48	&	13	&	2.5	&	0.4		&1.2		&0.027 \\            
 $^{67}$Cu  	&	1.4	& 	1.3	&	1.2	&	$>$25	&$>$75	&4.8\\            
 $^{83}$Br  	&	9.6	&	1.8	& 	0.3	& 	1.2 		& 3.7		& 0.29  \\            
 $^{90}$Y  	&	71	&	26	&	6.1	& 	0.4 		& 1.2		&  0.027 \\            
 $^{97}$Zr  	&	319 	&	143	&	102	& 	0.1  		&0.4		& 6.1\\  
 $^{131}$I  	&	12 	&	10	& 	9.9	&	$>$25	&$>$75 	&16\\            
 $^{133}$I  	&	44 	&	28	&	23 	&	2.2		&6.5 		& 23\\            
$^{153}$Sm  	&	3.0	&	0.7	&	0.3	&	4.7 		&14		& 5.0\\            
 $^{177}$Lu  	&	0.6	&	0.4	&	0.3	&	$>$25	&$>$75	& 1.6  \\            
$^{188}$Re  	&	55	&	18	&	4.3	&	0.5		&1.4		&    2.1 \\            
       \hline
       \end{tabular}
 \end{center}
\end{table}

\section*{Discussion}
This study compares the performances of a prototype probe optimized for $^{90}$Y detection on further radio-isotopes of interest. The results obtained are very important to guide on one side the choice of possible radio-tracers, on the other side the  design of future prototypes.

The results indicate that three classes of radioisotopes can be identified:
\begin{itemize}
\item those that can be used with the current prototype alternatively to $^{90}$Y: $^{31}$Si,$^{32}$P, $^{97}$Zr, and $^{188}$Re. Most of these isotopes are even more favorable than $^{90}$Y since they have an electron energy endpoint low enough to limit the background from nearby healthy tissues, but high enough  for the probe to be efficient. It is to be noted that lowering the minimum detectable energy would help anyhow, because it would allow detection also of partially infiltrated residues that are not completely at contact.
\item those that would profit significantly from a tuning of the detectable energy threshold: $^{83}$Br, $^{133}$I, and $^{153}$Sm.  It is though to be noted that, for SUV$\ge$4 and  TNR$>5$, $^{83}$Br and $^{133}$I yield significant signals. 
\item those that in addition to a reduction in detectable energy threshold would require also further photon suppression: $^{18}$F,$^{67}$Cu, $^{131}$I, and $^{177}$Lu.  Furthermore \fluoro has a low end-point that makes also the detection efficiency low. The design of a detector suited for them is very challenging. It is in particular to be stressed that in the case of $\beta^+$ emitters the large $\gamma$ background cannot be overcome without a dual-detector approach. 
\end{itemize}

As far as the medical personnel exposure is concerned, the effective dose rate is negligible for the pure $\beta^-$ emitters ($^{31}$Si,$^{32}$P, and $^{90}$Y), while for the rest of the radio-isotopes for which the current prototype could be used at least under some conditions, $DR_{med}$ is below $\approx 6\mu$Sv/hr. In any case the dose to the medical personnel is at least one order of magnitude smaller than if a $\beta^+$ emitter (like \fluoro) were used. $^{133}$I is again an exception and an efficient use would be possible only by  lowering the administered activity.

\section*{Conclusions}

RGS with $\beta^-$ decays can be performed also with radio-isotopes other than $^{90}$Y. This paper has established a first set of possible alternatives: $^{31}$Si,$^{32}$P, $^{97}$Zr, and $^{188}$Re. This  information is the starting point for the identification of new possible applications of the technique and for the development of new radio-tracers.

Finally, this study has identified other possible radio-isotope that could be useable if the minimum detectable energy could be optimized or if SUV and TNR conditions are particularly favourable ($^{83}$Br,  $^{133}$I and $^{153}$Sm) and it has set a procedure to establish whether a detector improvement is sufficient to broaden the range of possible applications of the RGS technique.

%
%

\section*{Acknowledgments}
We would like to thank G. Chiodi and L. Recchia of the electronics shop of the INFN Sez. di Roma for the hardware and software support in the realization of the probe prototypes. This research was funded in part by AIRCÐFondazione Cariplo TRIDEO Id. 17515  and  PRIN 2012 (prot. 2012CTAYSY).
\section*{Author Contributions}
G.B., C.M.T., E.S.C., V.B., R.D.,  M.M., A.R., M.T.  and S.M. designed and realized the experimental setups and took the data.
C.M.T. , F. C., R.F., E.S.C., and G.T. performed the simulations and the data analysis. 
A.C., I.F., D.R., and T.S. evaluated the relevance of the radio-isotopes from the chemical point of view while A.G. evaluated the potentialities from a Nuclear Medicine point of view. L.I. performed the radio-protection studies. Everybody contributed to the paper writing,  coordinated by R.F.

\section*{Competing Finantial Interests}
F.C.,  M.M.,  and R.F
are listed as inventors on an Italian patent application 
(RM2013A000053) 
entitled
``Utilizzo di radiazione $\beta^-$ per la identificazione intraoperatoria 
di residui tumorali e la corrispondente sonda di rivelazione" 
dealing with the implementation of an
intra-operative $\beta^-$ probe for radio-guided surgery
according to the results presented in this paper. 
The same authors are also inventors in the PCT patent application (PCT/IT2014/000025) entitled 
"Intraoperative detection of tumor residues using beta- radiation and corresponding probes "
covering the method and the instruments described in this paper.



\end{document}